# THE NIF INTEGRATED TIMING SYSTEM — DESIGN AND PERFORMANCE

R. A. Lerche, G. W. Coutts, L. J. Lagin, R. A. Nyholm, N. R. Sewall, R. D. Stever, J. D. Wiedwald,
Lawrence Livermore National Laboratory, PO Box 808 MS L-481, Livermore, CA 94550, USA
J. Larkin, Highland Technology, Inc., 320 Judah St., San Francisco, CA 94122, USA
S. Stein, R. Martin, Timing Solutions Corp., 5335 Sterling Dr, Ste B, Boulder, CO, 80301 USA

Abstract

The National Ignition Facility (NIF) will contain the world's most powerful laser. NIF requires more than 1500 precisely timed trigger pulses to control the timing of laser and diagnostic equipment. The Integrated Timing System applies new concepts to generate and deliver triggers at preprogrammed times to equipment throughout the laser and target areas of the facility. Trigger pulses during the last 2 seconds of a shot cycle are required to have a jitter of less than 20 ps (rms) and a wander of less than 100 ps (max). Also, the Timing System allows simultaneous, independent use by multiple clients by partitioning the system hardware into subsets that are controlled via independent software keys. The hardware necessary to implement the Integrated Timing System is commercially available.

## 1 INTRODUCTION

The National Ignition Facility (NIF) will include the world's most powerful laser (1.8 MJ) and a target irradiation area where millimeter-size targets containing isotopes of hydrogen will be compressed and heated to generate a fusion reaction. [1] Synchronization of NIF's 192 laser beams and laser and target diagnostics requires that ~1500 precisely timed trigger pulses be distributed to equipment throughout the 275,000-square-foot facility. The Integrated Timing System (ITS), a component of the NIF Integrated Computer Control System (ICCS) [2], provides these precision trigger signals.

Architecture of the ITS differs significantly from that used at most earlier pulsed laser facilities. In previous facilities, a single optical or electrical pulse was distributed over a tree structure, regenerated and redistributed (potentially several times) until it eventually triggered many preprogrammed delay generators. Such systems (1) restrict all users to a common, system-wide pulse rate; (2) result in progressively increasing jitter with the addition of active elements; and (3) when reconfigured, inevitably affect trigger channels where no change was intended.

The ITS uses fiber-optic communications and network technologies to distribute a time-encoded data stream throughout the facility; then generates independently controlled, real-time trigger pulses close to the device to be triggered. The NIF timing technique is an extension of that used at the OMEGA laser facility [3]. Precision timing of 20 to 100 ps is achieved during the last 2 seconds of a shot cycle. Real-time triggers are needed to synchronize laser oscillators, flash lamps, Pockels cells, and laser and target diagnostic instruments. Besides its programmable timing capability, the ITS allows simultaneous, independent use by multiple clients for setup and maintenance tasks. This paper describes the ITS architecture, its major components, and system performance.

## 2 ITS DESIGN

### 2.1 Architecture Overview

The ITS [4] consists of two major functional elements: a Facility Timing System (FTS) and a Local Timing System (LTS). The basic system architecture is shown in Figure 1. The FTS is an electro-optic system that generates and distributes a time-encoded, serial data pattern through a fiber-optic network to 14 zones located throughout the NIF. These zones contain eight-channel delay generator (DG) modules that comprise the LTS. The DGs receive and decode the data patterns and generate the critically timed electrical and optical triggers required by NIF hardware.

### 2.2 Facility Timing System (FTS)

The FTS includes a Master Clock, a Master Timing Transmitter (MTT), a Measurement System (MS), and a controller that are centrally located at the Facility Timing Node, and a Fiber-Optic Distribution Network.

The master clock provides stable 10-MHz and 1-Hz signals to the MTT, which synthesizes the signals into a standard communication frequency at 155.52-MHz (SONET OC-3). The MTT generates and transmits the serial data stream (SDS) that produces the master triggers and precision clocks needed by the remote DGs. The data encoded into the SDS consists of internally generated data (real-time clock, epochs, frame synchronization pattern (FSP), CRC, fill bits) and externally controlled data (trigger keys). Data are grouped into packets or frames (see Section 2.3). Data frame contents are converted into a bi-phase encoded SDS with a 77.76 Mbit/sec transfer rate. Conversion of the SDS from an electrical signal to an optical signal occurs in the MTT. The output data stream is split 16 ways to provide identical SDSs for up to 15 remote locations and a reference signal for the MS.

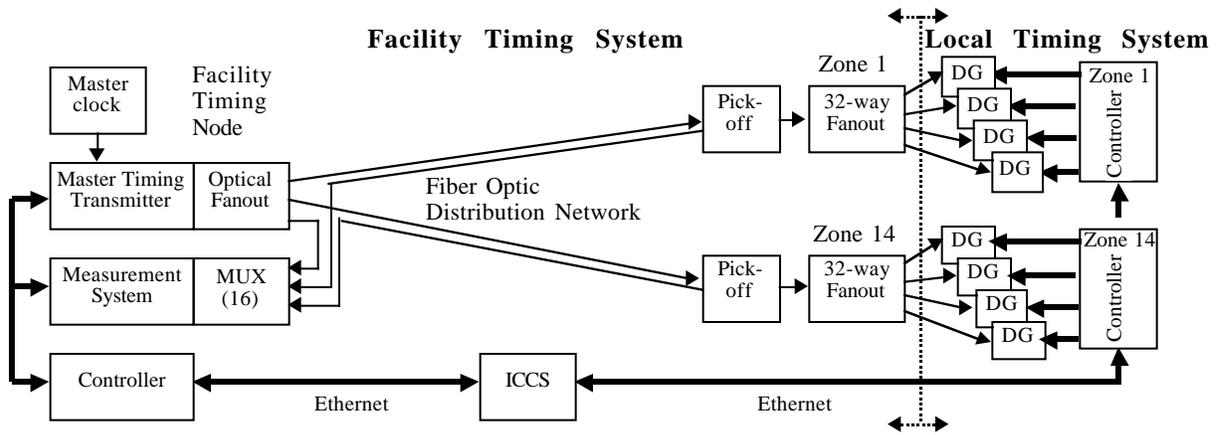

**Figure 1.** Block diagram showing major functional components of the NIF Integrated Timing System. MUX (multiplexer), DG (8-channel Delay Generator module).

The MS is co-located with the MTT. This unit receives a copy of the SDS from each local area fan-out and a reference copy of the data stream directly from the adjacent transmitter. These data streams are multiplexed to a decoder where the precise time of arrival is determined and passed to the controller. This provides the link transit time between the MTT and each local area fan-out and allows for time compensation at the DGs.

A controller module provides the communication interface between Facility Timing Node hardware and the remote ICCS computers. Programs running on the controller receive commands and data from the ICCS computers via an Ethernet connection and provide the user with necessary control and monitoring of the FTS.

Timing Solutions Corporation provides the MTT, MS, and the controller. These units are implemented separately, but use a modular hardware structure.

The Fiber Optic Distribution Network is comprised of standard communications hardware operating at 1550 nm. A cable containing two single-mode optical fibers connects the optical fan-out of the MTT with each of the remote zones of the LTS, which can be up to 500 meters away. One fiber delivers the encoded data stream as a high-level optical signal to the local area fan-outs; the other fiber returns a low-level replica of the data stream from each local area pick-off to the MS of the FTS. Signals on this return fiber verify link continuity and permit fiber transit times to be measured and monitored.

In each local area, the signal is optically split and can be delivered to up to 32 DG modules.

## 2.3 Serial Data Stream (SDS)

The SDS is central to the design of the ITS. It is formed in the MTT and is used to transfer serially encoded digital data and a clock signal to local area DGs (see Figure 2).

The SDS is assembled from internal and external data. It is then transmitted as sequential, contiguous, fixed-length data frames separated by a FSP. The frame rate is 24,000 frames per second. The SDS consists of static and dynamic data. Static data such as the FSP and fill bits (Fb) do not change from frame to frame.

Dynamic data that may change with each data frame include asynchronous and synchronous keys (AK and SK), epochs (E), real-time clock (RTC), and frame counter (FC). Data bits referred to as "keys" are under user control and change as directed. Asynchronous keys (2 bits) are updated near real time (within one frame) through the control interface or directly through external connectors. Synchronous keys (46 bits) which function as DG triggers are updated through the control interface once per second. Synchronous keys are used to partition the timing system among various users. Epochs (8 bits) are periodic events that occur for a single frame at specific rates between 960 and 0.2 Hz. A single bit transmitted at a subharmonic of the frame rate represents each epoch. RTC and FC data (72 bits) are time-dependent bits that are initialized through the control interface. Since each frame occurs at a unique time, at least one bit of this data will change with each frame.

Data are bi-phase encoded, i.e., one data bit is encoded for every two clock periods. Bi-phase encoding allows a stable clock whose phase is nearly independent of the data to be recovered from the SDS and also permits arbitrarily long runs of either "ones" or "zeros".

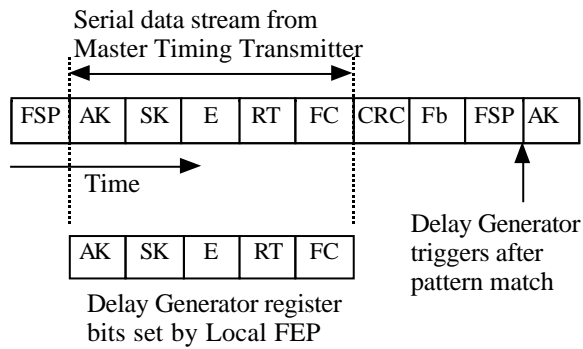

**Figure 2.** Serial data stream. Delay generator is triggered by first FSP after a data frame match.

## 2.4 Programmable Delay Generator (DG)

Highland Technology, Inc. manufactures the programmable DG. Their model V880 has a VME format with eight independent timing channels, one reference channel, and one monitor channels per module. The monitor channel can be programmed to monitor one of the eight output channels. DGs having either optical or electrical outputs are available. When operated in local mode, the unit functions like a conventional DG. Three unique features have been added to these modules:

1. The clock is recovered from the SDS. This provides a clock in each DG whose frequency is common to all DGs throughout the NIF facility and whose phase is fixed to the received data stream. The clock is used for delay countdown and also serves as a synchronizing edge for triggers.

2. A trigger is initiated in a DG channel when a pattern match occurs between a transmitted data frame and data preprogrammed into the DG data register. This allows DG channels to be grouped using key and epoch bits. At the same time, users of the other DG channels can operate independently by using other keys and epochs. This is especially useful during activation and maintenance.

3. A successful pattern match causes a trigger delay to be initiated upon the next FSP. Synchronizing the trigger to the FSP simplifies DG implementation and improves performance by eliminating the need for a start synchronizer circuit. This removes a major source of timing jitter found in many conventional DGs.

Major contributors to long-term wander are temperature and SDS power variation. The DGs have built-in compensation for both temperature and optical power variations that allow them to meet NIF performance requirements.

## 3 ITS PERFORMANCE

The ITS has been tested at the component and at the system level. All hardware was component-level tested prior to performing integrated system tests. The MTT and MS were evaluated via extensive acceptance testing at the manufacturer. The MS, however, was not included in initial integrated system tests as acceptance tests verified its stability to be better than 10 ps for 5 days. DGs were acceptance tested at the manufacturer before shipping to LLNL and system testing.

For integrated system tests, the MTT SDS was connected to 19 V880 DGs via fiber optic splitters. Thirty-two outputs from 13 of the DGs were monitored using an array of multiplexed RF switches and one precision Wavecrest time interval meter (TIM). Each DG was programmed for a 350-$\mu$second delay. At 10-minute intervals, a computer switched the RF MUX through the 32 DG outputs, thereby routing them to the "stop" input of the TIM. For each DG channel, the TIM acquired 1000 DG output triggers. These measurements were reduced to two data values: an average delay time and peak-to-peak jitter of the delay time.

This measurement process was repeated every ten minutes for seven days while the environmental conditions and key DG module parameters were monitored and archived. After seven days, the recorded data were analyzed to determine wander and jitter performance.

Maximum wander over the seven-day test was 52 ps for one channel, while the other 31 channels had less than 32 ps. Jitter for all channels was <7 ps rms. The integrated system tests demonstrated that the ITS meets its seven-day wander and short-term jitter requirements of <100 ps (maximum) and <20 ps (rms), respectively.